\documentclass[aps,prl,twocolumn]{revtex4-2}

\usepackage{blindtext,amssymb,gensymb,siunitx,textgreek,textcomp,mathtools,bm,soul}
\usepackage{wasysym} 
\usepackage[version=3]{mhchem} 

\usepackage[utf8]{inputenc}
\usepackage[T1]{fontenc}
\usepackage [english]{babel}
\usepackage [autostyle, english = american]{csquotes}
\MakeOuterQuote{"}
\usepackage{subfiles}
\usepackage{fancyvrb}
\usepackage{braket}

\usepackage[margin=0.625in]{geometry}

\usepackage{caption,graphicx}
\usepackage{tablefootnote}
\usepackage{booktabs}
\usepackage[table,xcdraw]{xcolor}
\usepackage[hidelinks]{hyperref}
\usepackage{cleveref,xr}

\AtBeginDocument{%
  \LetLtxMacro\hyperrefautoref\autoref
  \LetLtxMacro\autoref\firstboldautoref
}
\makeatletter
\DeclareRobustCommand\firstboldautoref{\@firstboldautoref}
\def\@firstboldautoref#1#{%
  \def\fb@autoref@star{#1}%
  \fb@autoref
}
\def\fb@autoref#1{%
  \ifcsname boldautoref@#1\endcsname
    \expandafter\hyperrefautoref\fb@autoref@star{#1}%
  \else
    \global\expandafter\let\csname boldautoref@#1\endcsname\@empty
    \textbf{\expandafter\hyperrefautoref\fb@autoref@star{#1}}%
  \fi
}
\makeatother

\begin{document}

\title{Designing transparent conductors using forbidden optical transitions}
\author{Rachel Woods-Robinson\textsuperscript{1,2}}
\author{Yihuang Xiong\textsuperscript{3}}
\author{Jimmy-Xuan Shen\textsuperscript{4}}
\author{Nicholas Winner\textsuperscript{1,6}}
\author{Matthew K. Horton\textsuperscript{1}}
\author{Mark Asta\textsuperscript{1,6}}
\author{Alex M. Ganose\textsuperscript{5}}
\author{Geoffroy Hautier\textsuperscript{3}}
\author{Kristin A. Persson\textsuperscript{6,7}}

\affiliation{\\ \textsuperscript{1}Materials Sciences Division, Lawrence Berkeley National Laboratory, Berkeley, CA, USA, \textsuperscript{2}Applied Science and Technology Graduate Group, University of California at Berkeley, Berkeley, CA, USA,
\textsuperscript{3}Thayer School of Engineering, Dartmouth College, 14 Engineering Dr, Hanover, NH, USA,
\textsuperscript{4}Lawrence Livermore National Laboratory, Livermore, CA, USA,
\textsuperscript{5}Department of Chemistry, Molecular Sciences Research Hub, White City Campus, Imperial College London, Wood Lane, London, UK,
\textsuperscript{6}Department of Materials Science and Engineering, University of California at Berkeley, Berkeley, CA, USA,
\textsuperscript{7}Molecular Foundry Division, Lawrence Berkeley National Laboratory, Berkeley, CA, USA, \
}

\date{\today}

\begin{abstract}

Many semiconductors present weak or forbidden transitions at their fundamental band gaps, inducing a widened region of transparency. This occurs in high-performing n-type transparent conductors (TCs) such as Sn-doped \ce{In2O3} (ITO), however thus far the presence of forbidden transitions has been neglected in searches for new p-type TCs. To address this, we first compute high-throughput absorption spectra across $\sim$18,000 semiconductors, showing that over half exhibit forbidden or weak optical transitions at their band edges. Next, we demonstrate that compounds with highly localized band edge states are more likely to present forbidden transitions. Lastly, we search this set for p-type and n-type TCs with forbidden or weak transitions. Defect calculations yield unexplored TC candidates such as ambipolar \ce{BeSiP2}, \ce{Zr2SN2} and \ce{KSe}, p-type \ce{BAs}, \ce{Au2S}, and \ce{AuCl}, and n-type \ce{Ba2InGaO5}, \ce{GaSbO4}, and \ce{KSbO3}, among others. We share our data set via the MPContribs platform, and we recommend that future screenings for optical properties use metrics representative of absorption features rather than band gap alone.



\end{abstract}

\maketitle



\section{Introduction}

\begin{figure*}
    \centering
    \includegraphics[width=\textwidth]{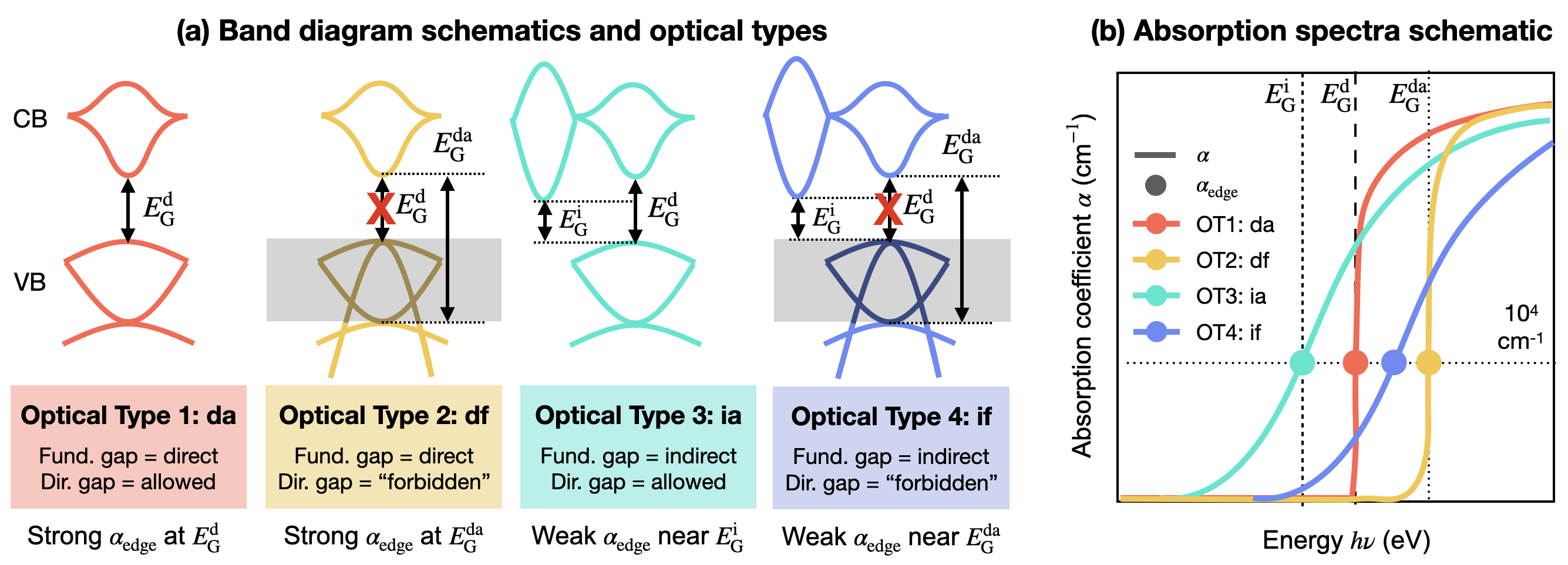}
    \caption[Band schematics and absorption spectra of various semiconductor material optical types.]{(a) Band schematics and (b) cartoon of the resulting absorption spectra of the four optical types (OTs) in semiconductor materials. Band schematics on left are inspired by Yu and Zunger.\cite{yu2012identification} The grey regions in OT2 and OT4 correspond to the forbidden region where transitions do not occur. ``Fund. gap'' stands for fundamental electronic band gap and ``dir. gap'' stands for direct band gap.}
    \label{fig:optical-types}
\end{figure*}

It is often assumed in semiconductors that a strong absorption onset occurs at the direct fundamental band gap. This is indeed the case for many materials, however some materials have forbidden transitions at their band edges such that the onset of their absorption edge occurs at higher energies than their direct gap. Four scenarios of absorption in semiconductors are depicted schematically in \autoref{fig:optical-types}, following the optical type (OT) classification as outlined by Yu and Zunger\cite{yu2012identification} for four hypothetical materials with similar band structures. In OT1 the fundamental band gap $E_\mathrm{G}$ is direct and allowed (``da''), in OT2 $E_\mathrm{G}$ is direct but forbidden (``df''), in OT3 $E_\mathrm{G}$ is indirect and the direct gap is allowed (``ia''), and in OT4 $E_\mathrm{G}$ is indirect and the direct gap is forbidden (``if'').

The presence of forbidden optical transitions can be detrimental in certain applications (e.g., LEDs, solar cell absorbers), however for others it may present a useful design criteria. In this study we focus on transparent conductors (TCs) --- materials combining wide optical transparency with high mobility and doping --- which require weak absorption within a given range of wavelengths (usually within the visible) such that forbidden transitions could be advantageous to increase transparency. In fact, many of the high-performing, commercially-available n-type transparent conducting oxides (TCOs) have dipole forbidden transitions at their band edges that induce this behavior. A notable example occurs in the most common TCO, n-type Sn-doped \ce{In2O3} (ITO), with weak absorption in the upper-most 0.8 eV of the valence band (VB), allowing for an increased transparency in addition to the increase from the Burstein-Moss effect.\cite{walsh2008nature} Other wide-gap oxide materials with reported forbidden transitions include \ce{SnO2} and F-doped \ce{SnO2} (FTO),\cite{summitt1964ultraviolet, frohlich1978band} spinels \ce{SnZn2O4}, \ce{SnCd2O4} and \ce{CdIn2O4},\cite{segev2005structure} \ce{Tl2O3},\cite{kehoe2011nature} and \ce{TiO2}.\cite{tang1994optical} Additionally, dipole-forbidden transitions have been reported in Cu-based p-type TCs including delafossites \ce{CuAlO2}, \ce{CuGaO2}, and \ce{CuInO2}, as well as cuprite \ce{Cu2O}.\cite{nie2002bipolar}

Meanwhile, it is of considerable interest to identify new high-performing p-type TC for applications in photovoltaics and beyond. Over the past decade, high-throughput screening studies have proposed several n-type or p-type TC candidates such as \ce{ZnSb2O6}, \ce{ZrOS}, \ce{BP}, \ce{Ba2BiTaO6}, and \ce{CaTe}\cite{hautier2013identification, hautier2014does, bhatia2015highmobility, ha2018computationallydriven, brunin2019transparent}. Experimental confirmation of exceptional properties has been demonstrated in some of those computationally-identified materials such as the p-type \ce{Ba2BiTaO6} and n-type \ce{ZnSb2O6},\cite{bhatia2015highmobility, jackson2022computational} but still no predicted p-type TC has experimentally-confirmed properties on par with n-type ITO. Most high-throughput screenings for TCs to date assume wide electronic band gap or direct band gap as a proxy for transparency.\cite{hautier2013identification,sarmadian2016easily, williamson2016engineering, shi2017highthroughput, 
varley2017highthroughput, ha2018computationallydriven} This assumption does not consider whether associated optical transitions are actually allowed or strong, thus overlooking materials with a small fundamental band gap but a wide absorption edge which could enable optical transparency. We note that several screenings for solar absorbers have explicitly considered forbidden transitions,\cite{yu2012identification, meng2017parity, fabini2019candidate} \textit{excluding} materials with forbidden edges to design for a sharp absorption onset; in contrast, a screening for TCs would \textit{include} such materials.

\begin{figure*}
    \centering
    \includegraphics[width=0.75\textwidth]{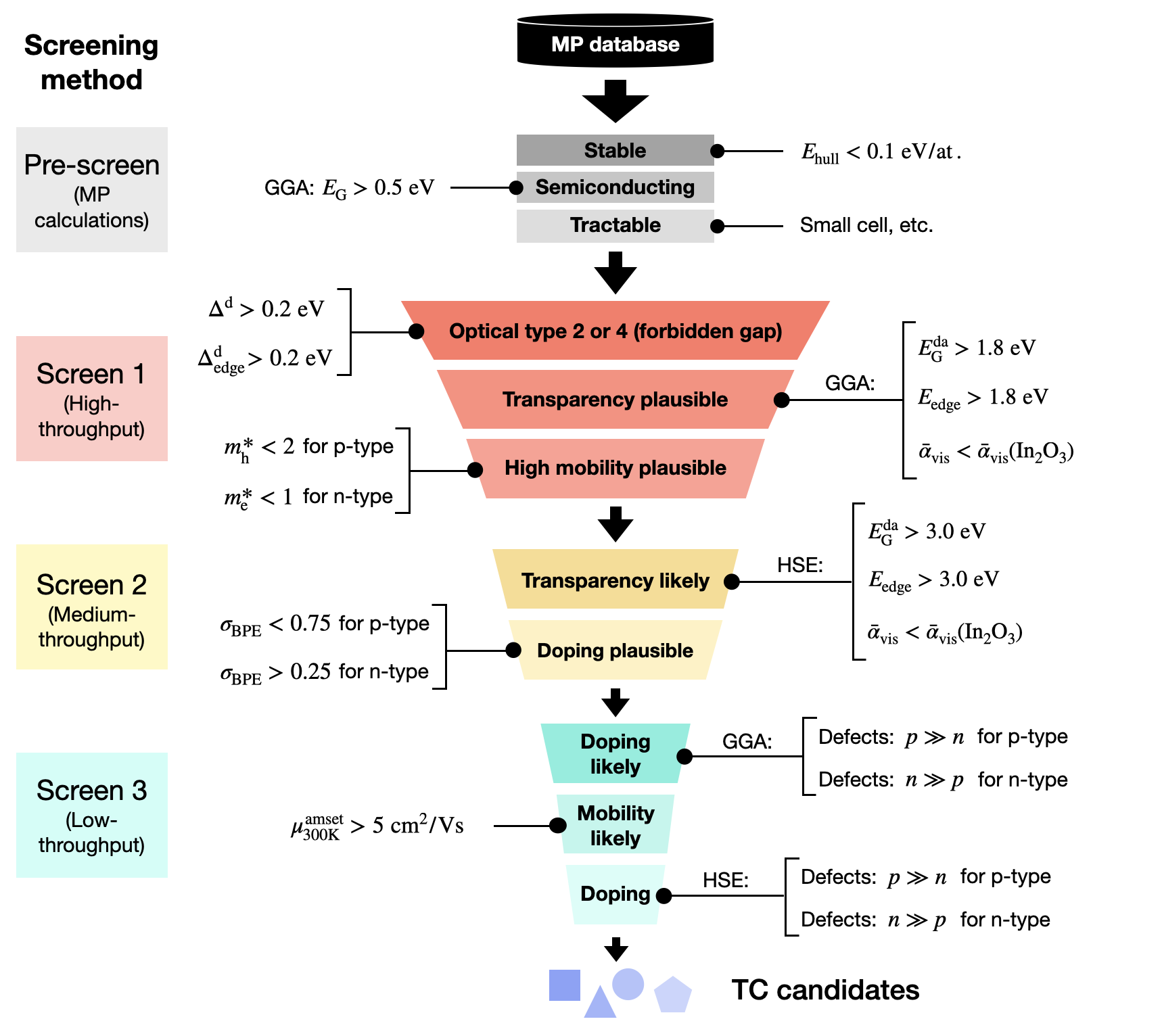}
    \caption{(a) The screening method for TCs pursued in this paper, focusing on compounds from the MP database with forbidden optical transitions. The targeted property is listed in the screen, and the descriptor and cutoff value are given on the right-hand side. Note that these descriptors are computed with both GGA (for screen 1) and HSE06 (for screen 2) functionals, and are described in more detail in the manuscript.}
    \label{fig:screening-method}
\end{figure*}

Therefore, in this work we leverage forbidden optical transitions at band edges (referred to hereafter as simply ``forbidden transitions'') to improve high-throughput searches for TCs. First, we benchmark and compute optical absorption edges for $\sim$18,000 inorganic compounds in the Materials Project (MP) database, and classify optical types across MP to assess whether the fundamental gaps are optically allowed or forbidden. We show that over half of the selected semiconductors in MP exhibit a weak absorption edge, and that, in special cases involving transitions between localized states, the presence of forbidden transitions can be explained by orbital character. With this data, we introduce a series of high-throughput descriptors for p-type TCs to estimate the direct allowed band gap (often referred to in the literature as the ``optical gap''), absorption edge onset, and average absorption spectra in the visible spectrum. Using these descriptors, we perform a high-throughput screening (as outlined in \autoref{fig:screening-method}) for promising p-type and n-type TCs with disperse band edges that may be transparent in the visible regime. Such compounds have low fundamental band gaps, and therefore may have previously been overlooked. To assess dopability and mobility for materials with good computed optical properties, we perform defect formation energy calculations and compute transport properties for the most promising candidates. We highlight some ambipolar TC candidates including \ce{BeSiP2}, p-type TC candidates including boron \ce{BAs}, and n-type TC candidates including barium indium gallium oxide (\ce{Ba2InGaO5}), and share our data for further exploration.


\section{Materials and methods}


Density functional theory (DFT) calculations were performed using the projector augmented wave (PAW) method\cite{blochl1994projector, kresse1999ultrasoft} as implemented in the Vienna \textit{Ab Initio} Simulation Package (VASP)\cite{kresse1993initio, kresse1996efficient}, first within the Purdue-Berke-Ernzerhof (PBE) Generalized Gradient Approximation (GGA) formulation of the exchange-correlation functional.\cite{perdew1996generalized} Cutoff, convergence, and correction criteria have been benchmarked and are used throughout the MP infrastructure, as described elsewhere.\cite{ong2013python, jain2013commentary} Effective mass ($m^*$) was computed from GGA calculations using the \texttt{BoltzTraP2} package,\cite{madsen2018boltztrap2} assuming dopings of $10^{18}$ cm\textsuperscript{-3} as described in the Supplementary Materials (SM). The HSE06 screened hybrid functional\cite{heyd2003hybrid} was used to calculate gap corrections and apply scissor shifts in ``screen 2.'' Branch point energy (BPE) was computed from GGA calculations with an HSE gap correction; BPE ratio range $\sigma_\mathrm{BPE}$ was computed by varying number of valence bands ($N_\mathrm{VB}$) and number of conduction bands ($N_\mathrm{CB}$) from $N_\mathrm{VB}$:$N_\mathrm{CB}$=2:4 to $N_\mathrm{CB}$:$N_\mathrm{CB}$=8:4, with details described elsewhere.\cite{woods-robinson2018assessing} The site-projected wave function character of orbitals at the band edges were assessed to compute the inverse participation ratios (IPRs) and the orbital overlaps (see SM).

Optical absorption coefficients were calculated with VASP using the independent-particle approximation (IPA). Using the IPA, the dielectric matrix elements are calculated using a k-point reciprocal density of 1,000 Å\textsuperscript{-3}, which we have benchmarked and optimized for high-throughput screenings $E_\mathrm{edge}$ (for optimization of precision in the extended absorption spectrum, see Yang et al.\cite{yang2022high}). Cutoff for a transition to be considered ``allowed'' was selected following convention from Fabini et al.\cite{fabini2019candidate} Details and calculation parameters for this method are reported in the SM.

Focusing on compounds that are likely to be synthesizable and are tractable for further defect calculations, we ``pre-screen'' (see \autoref{fig:screening-method}) the MP database using a series of filters. We include compounds in which the MP computed GGA fundamental band gap ($E_\mathrm{G}$) is greater than 0.5 eV and the energy above convex hull ($E_\mathrm{hull}$) is less that 0.1 eV/atom.\cite{aykol2018thermodynamic, sun2016thermodynamic} Large compounds were filtered out with more than 5 elements or more than 12 symmetrically inequivalent sites (see \texttt{pymatgen.symmetry.analyzer}). Compounds with heavy elements ($Z$ > 82) and f-block elements are also filtered out (except for La). GGA absorption spectra of $\sim$800 MP compounds from Fabini et al.'s search for PV absorber materials are publicly available on MPContribs and included in our set.\cite{fabini2019candidate, huck2016user}

For compounds that emerge from ``screen 2,'' defect formation energy calculations are performed using the \texttt{pycdt} package,\cite{Broberg2018defect}. Hybrid density functional theory calculations of defect formation energies are performed using the CP2K software package and HSE06 functional.\cite{hutter2014cp2k, kuhne2020cp2k,heyd2003hybrid} Charge-carrier mobility is calculated using the \textit{ab initio} scattering and transport package (\texttt{amset}),\cite{ganose2021efficient} which solves the linearized Boltzmann transport equation under the constant relaxation time approximation. Details for each of these methods are described in the SM.

\begin{figure*}
    \centering
    
    \includegraphics[width=\textwidth]{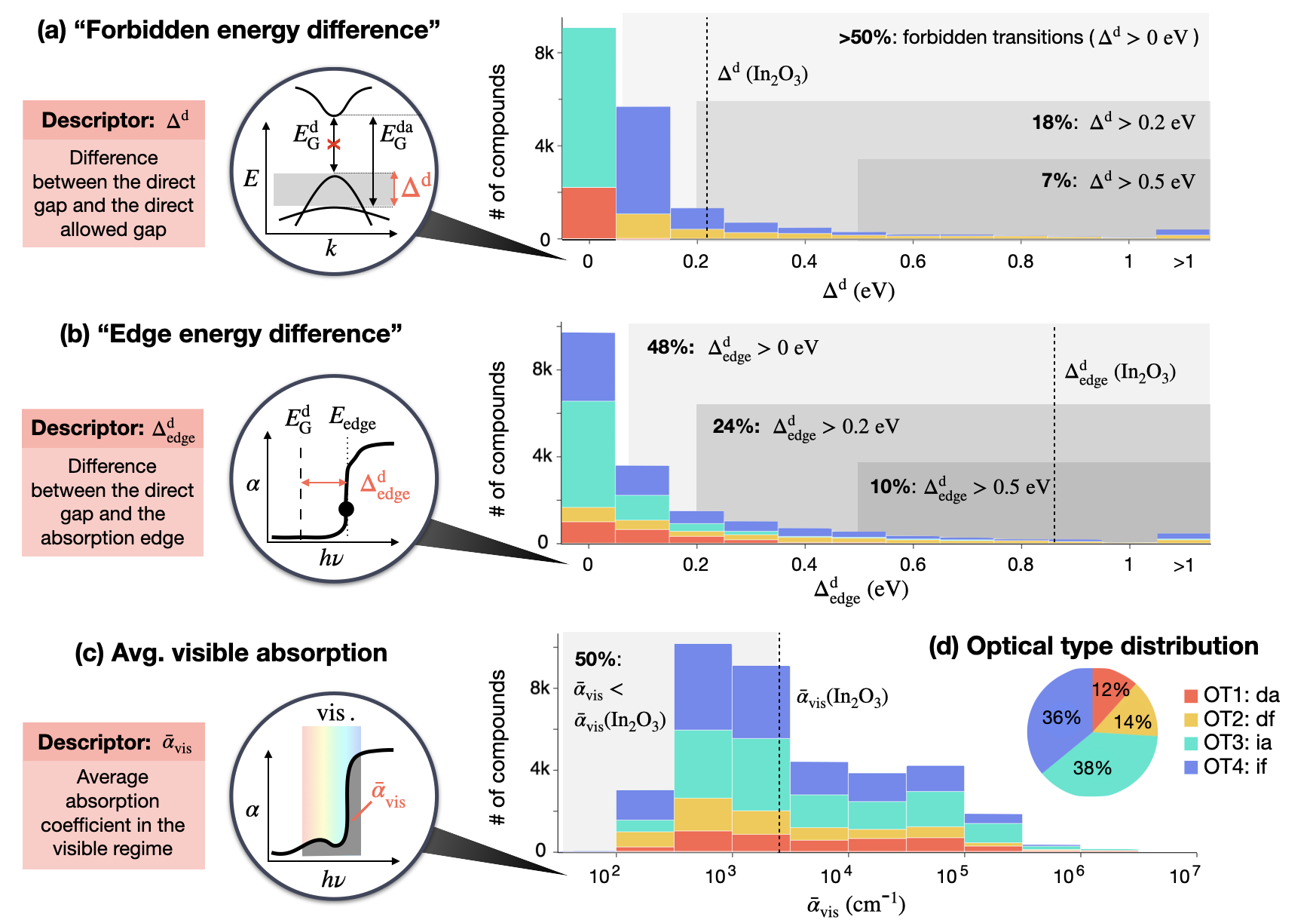}
    \caption{(Left) Schematics highlighting new optical screening descriptors: (a) ``forbidden energy difference'' $\Delta^\mathrm{d}$, (b) edge energy difference $\Delta^\mathrm{d}_\mathrm{edge}$, and (c) average absorption in the visible regime $\bar{\alpha}_\mathrm{vis}$. (Right) Histograms of these three optical descriptors are reported across the set of 18,000 semiconductors. Corresponding values for \ce{In2O3}, the best performing n-type TC, are denoted for reference. (d) Optical type (OT) distribution, showing over half of this set has a forbidden direct optical transition, 18\% exhibit $\Delta^\mathrm{d}$ > 0.2 eV, and 7\% exhibit $\Delta^\mathrm{d}$ > 0.5 eV.}
    \label{fig:forbidden-stats}
\end{figure*}

\section{Results}

\subsection{Forbidden or weak transitions are common}

As a result of the pre-screening, we obtain a data set of $\sim$18,000 semiconductors compounds for which optical absorption spectra and descriptors are assembled. Statistics and corresponding descriptors are summarized in \autoref{fig:forbidden-stats}, grouped by optical type. We first assess the distribution of optical types (OTs) and forbidden optical transitions across the set. To our knowledge, this has not been assessed across known semiconductor materials, except for the several hundred from Fabini et al.\cite{fabini2019candidate} \autoref{fig:forbidden-stats}(a) plots a histogram of the descriptor ``forbidden energy difference'' $\Delta^\mathrm{d}$, defined as:

\begin{equation}
    \Delta^\mathrm{d} = E_\mathrm{G}^\mathrm{da} - E_\mathrm{G}^\mathrm{d},
    \label{eq:delta-d}
\end{equation}

    

\noindent where the direct allowed band gap $E_\mathrm{G}^\mathrm{da}$ is defined as the energy at which dipole transition matrix elements become significant (adopting what constitutes as "significant" from the literature;\cite{fabini2019candidate} see Supplementary Materials, SM). We demonstrate that nearly 50\% of compounds have forbidden transitions (i.e., $\Delta^\mathrm{d}$ > 0 eV) at the band edges. A large subset show a strong impact of forbidden transitions, $\sim$18\% with $\Delta^\mathrm{d}$ > 0.2 eV and 7\% with $\Delta^\mathrm{d}$ > 0.5 eV. It is observed that OT3 (indirect gap, allowed direct transition) is the most common optical type, followed closely by OT4 (indirect fundamental gap, forbidden direct transition).

\autoref{fig:forbidden-stats}(b) reports the distribution of the ``edge energy difference'' $\Delta^\mathrm{d}_\mathrm{edge}$, defined as:

\begin{equation}
    \Delta^\mathrm{d}_\mathrm{edge} = E_\mathrm{edge} - E_\mathrm{G}^\mathrm{d},
    \label{eq:delta-d-edge}
\end{equation}

\noindent where the absorption edge energy ($E_\mathrm{edge}$) is defined as the energy at which the GGA IPA absorption coefficient exceeds 10\textsuperscript{4} cm\textsuperscript{-1} (see SM). Some materials may have transitions at the band edges that are ``allowed'' but are only weakly absorbing; in these cases $E_\mathrm{edge}$ can provide a better metric than $E_\mathrm{G}^\mathrm{da}$ for where the strong edge onset actually occurs. For example, in \ce{In2O3} (dashed lines) our computed $\Delta^\mathrm{d}_\mathrm{edge}$ of 0.68 eV corresponds better to the literature-reported $\Delta^\mathrm{d}$ than our computed GGA $\Delta^\mathrm{d}$ of 0.22 eV. Third, in \autoref{fig:forbidden-stats}(c) we plot the distribution of the average absorption in the visible, $\bar{\alpha}_\mathrm{vis}$, defined as:

\begin{equation}
    \bar{\alpha}_\mathrm{vis} = \int_{\nu _\mathrm{vis}^\mathrm{min}}^{\nu _\mathrm{vis}^\mathrm{max}} \alpha(h \nu) ,
    \label{eq:alpha-vis}
\end{equation}

\noindent i.e., the integral of the absorption spectra across the visible regime. Since GGA Kohn-Sham gap underestimates fundamental band gap, we define the limits of the integral for GGA calculations using an empirical gap correction from Morales et al. (see SM).\cite{morales2017empirical} It is observed that in more than 50\% of compounds, $\bar{\alpha}_\mathrm{vis}$ is less than that of \ce{In2O3}. Of interest to this study are the set of compounds with a significantly widened absorption edge due to forbidden transitions and correspondingly low absorption coefficients. In particular, we are interested in materials in which forbidden transitions raise the absorption edge outside of the visible spectrum and lead to optical transparency. However we note this data set may be valuable for other investigations as well.


\subsection{Underlying chemical trends}

Forbidden and weak optical transitions in semiconductors can arise from a variety of physical, structural, and chemical phenomena. Inversion symmetry at the band extrema can induce parity forbidden dipole transitions,\cite{yu2005fundamentals} and a series of selection rules determine whether transitions can occur between states of different irreducible representation. Parity-forbidden transitions are invoked, e.g., for \ce{In2O3}, among other materials, to explain in part why experimental optical band gaps exceed the fundamental gap. According to Fermi's Golden Rule, if symmetry allows, transitions between localized states composed of similar chemical orbitals (i.e., with significant $\braket{\psi_\mathrm{i}|r|\psi_\mathrm{f}}$, which scales with overlap $\braket{\psi_\mathrm{i}|\psi_\mathrm{f}}$) have weak dipole transition matrix elements.\cite{yu2005fundamentals} However, due to the delocalized nature of wavefunctions in solids, understanding the mechanisms behind forbidden and allowed transitions is less straightforward in semiconductors than in molecules with discrete, localized states. 

\begin{figure*}
    \centering
    \includegraphics[width=\textwidth]{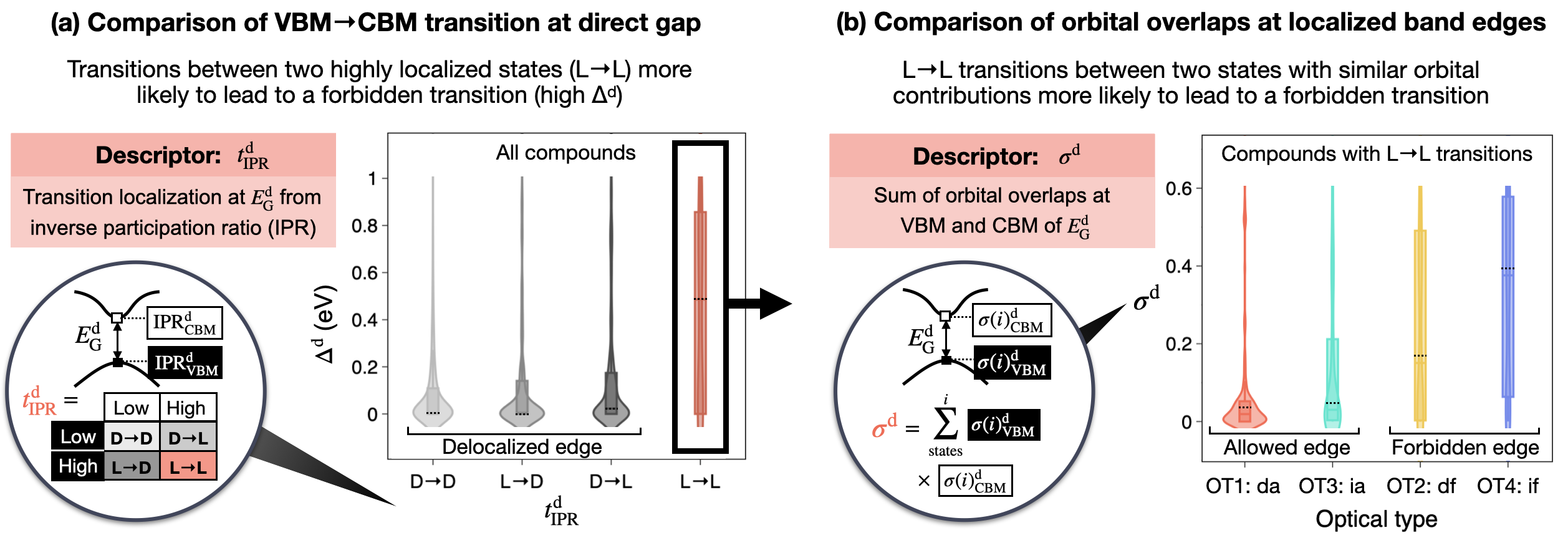}
    \caption{A schematic of band edge orbital descriptors and violin plots correlating each descriptor to the forbidden transitions data set for (a) $t_\mathrm{IPR}^\mathrm{d}$, the transition localization at the VBM and CBM from the inverse participation ratio (IPR) and (b) $\Sigma \sigma^\mathrm{d}$, the orbital overlap of the VBM and CBM states at $E_\mathrm{G}^\mathrm{d}$. In (a), ``D'' indicates a delocalized state and ``L'' indicates a localized state, and the scenario in which there are transitions from a highly localized state at the VBM to another highly localized state at the CBM (L$\rightarrow$L) is highlighted. In (b), only compounds with L$\rightarrow$L are plotted, and the violin plot shows the distribution of $\Sigma \sigma^\mathrm{d}$ across different optical types (note that OT2 and OT4 indicate forbidden transitions, i.e., $\Delta^\mathrm{d}$ > 0 eV.) In both violin plots, mean values of each distribution are reported with dashed black lines.}
    \label{fig:localization-and-orbitals}
\end{figure*}

Here, we explore whether the nature of forbidden transitions between the direct band edges can be correlated with two simple orbital-based descriptors, described in \autoref{fig:localization-and-orbitals}:

\begin{enumerate}

    \item \textbf{Inverse participation ratio of the direct VBM and CBM states,} $\bm{t_\mathrm{IPR}^\mathrm{d}}$: We consider the inverse participation ratio (IPR) across all compounds as a proxy for localization of states at the band edges (a high IPR indicates strong localization). As shown in \autoref{fig:localization-and-orbitals}(a), ``D'' indicates a delocalized state and ``L'' indicates a localized state, and values of descriptor $t_\mathrm{IPR}^\mathrm{d}$ are assigned as shown in the call-out circle (e.g., $t_\mathrm{IPR}^\mathrm{d}$ = ``L$\rightarrow$L'' indicates a transition from a strongly localized VBM to a strongly localized CBM).
    
    \item \textbf{Orbital overlap of the direct VBM and CBM states,} $\bm{\sigma^\mathrm{d}}$: For each compound, we consider the dominant contributors to the density of states at the direct VBM and CBM, $\sigma(l,m)^\mathrm{d}$ ($l$ is angular momentum quantum number $s$, $p$, $d$, or $f$ for each element, and $m$ is magnetic quantum number; see SM for details). With this data, we compute a descriptor $\sigma^\mathrm{d}$ (i.e., $\braket{\psi_\mathrm{i}|r|\psi_\mathrm{f}}$) to describe the similarity of CB edge and VB edge orbital contributions. This is depicted in the call-out circle in \autoref{fig:localization-and-orbitals}(b) as:

    \begin{equation}        
        \sigma^\mathrm{d} = \sum_\mathrm{l,m}\sigma(l,m)_\mathrm{VBM}^\mathrm{d}\sigma(l,m)_\mathrm{CBM}^\mathrm{d}
        \label{eq:orbitals}
    \end{equation}

    \noindent We will refer to this descriptor as "orbital overlap" in this paper.

    
\end{enumerate}

\begin{figure*}
    \centering
\includegraphics[width=\textwidth]{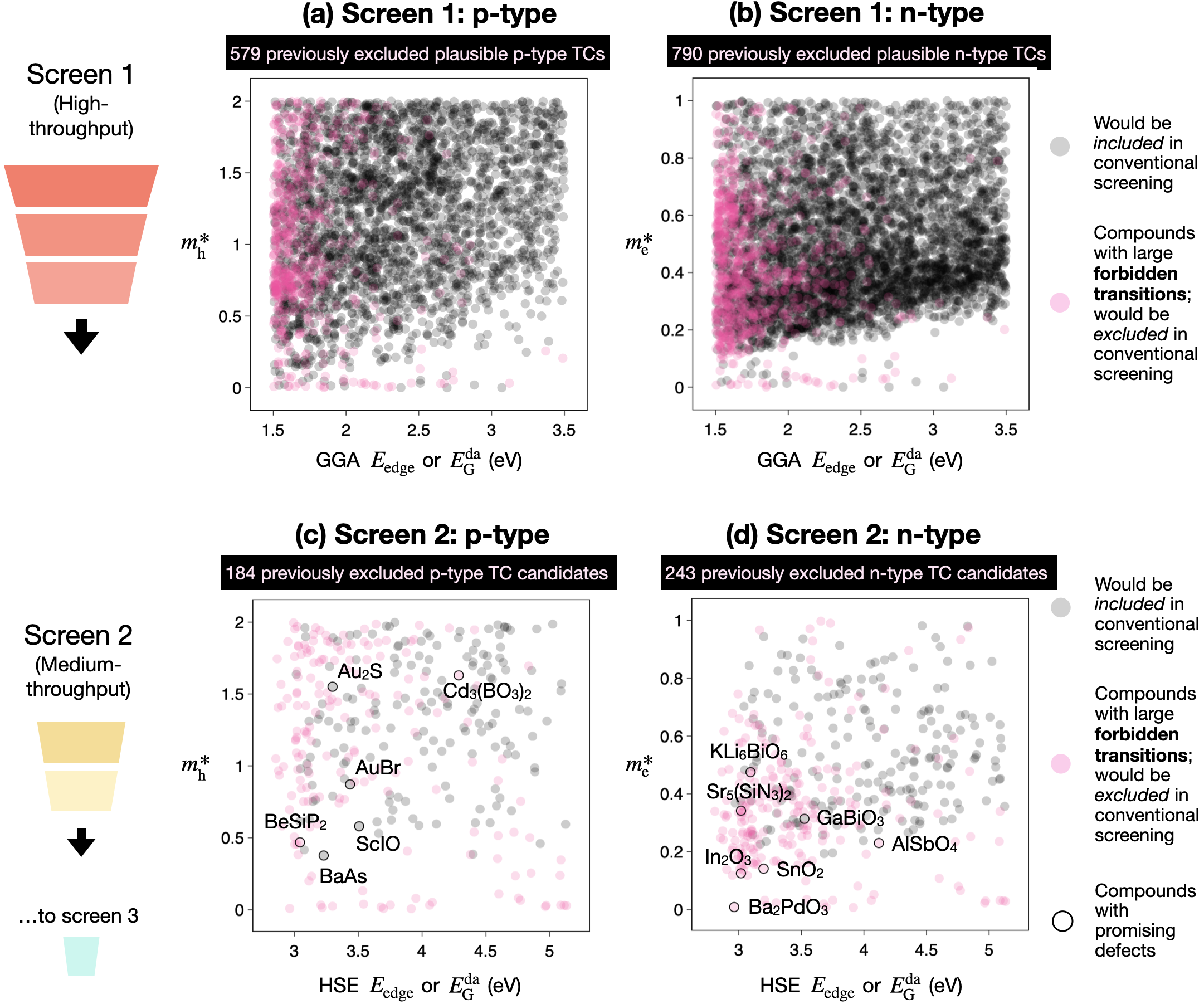}
    \caption{Screen 1 outputs using a GGA functional for (a) plausible p-type TC candidates (focusing on $m_\mathrm{h}^*$ < 2) and (b) plausible n-type TC candidates (focusing on $m_\mathrm{e}^*$ < 1). Screen 2 outputs using a HSE functional for (c) p-type TC candidates (focusing on $m_\mathrm{h}^*$ < 2) and (d) plausible n-type TC candidates (focusing on $m_\mathrm{e}^*$ < 1). In all four plots, computed $m^*$ is plotted as a function of either $E_\mathrm{G}^\mathrm{da}$ or $E_\mathrm{edge}$, depending on which value is higher, to reflect the screening procedure. Marker color denotes whether compounds would have been included (blue) or excluded (red) from a conventional screening in which the allowed or forbidden nature of the direct gap is not considered. Candidates emerging from screen 3 are highlighted in (c) and (d).}
    \label{fig:E-vs-meff}
\end{figure*}

Basic trends between these descriptors and the forbidden nature of the gap are summarized in \autoref{fig:localization-and-orbitals}. In the violin plot in (a), it is shown that in compounds in which states at both band edges are delocalized (D$\rightarrow$D), the forbidden energy difference $\Delta^\mathrm{d}$ is low with a mean close to zero. In compounds where at least one band edge is delocalized (D$\rightarrow$L and L$\rightarrow$D), the average $\Delta^\mathrm{d}$ increases slightly. However, compounds where both edges are highly localized (L$\rightarrow$L) are significantly more likely to have wide forbidden transitions; the average $\Delta^\mathrm{d}$ across such compounds is $\sim$0.5 eV, and quartiles range from 0.1--0.6 eV. Therefore, transitions between two highly localized states are likely to lead to a forbidden transition.

To inspect cases with localized transitions in more detail, we compute the orbital overlap $\sigma^\mathrm{d}$ for the L$\rightarrow$L subset from (a), and report the distribution of $\sigma^\mathrm{d}$ across the four optical types in the violin plot in \autoref{fig:localization-and-orbitals}(b). It is observed that, systematically, compounds with allowed edges (OT1 and OT3) have significantly lower orbital overlaps than compounds with forbidden edges (OT2 and OT4). This is consistent with Fermi's Golden Rule: low transition dipole matrix elements (i.e., forbidden or weak transitions) should occur between localized states with similar orbital contributions. We note that a weaker trend occurs when $\sigma^\mathrm{d}$ is plotted across all compounds (see SM), mostly likely because the selection rules from Fermi's Golden Rule break down in transitions between delocalized states.

Therefore, in cases with highly localized band edges, $\sigma^\mathrm{d}$ is a useful predictor for the origin of forbidden transitions. However, these L$\rightarrow$L cases are only a small subset ($\sim$10\%) of compounds in which we predict forbidden transitions, and there are other factors that arise for example due to the delocalized or hybridized nature of edge states. Ultimately, due to the relatively cheap computational cost of high-throughput IPA calculations and the variety of mechanisms contributing to optical transition matrix elements, we recommend further DFT calculations at this stage.



\subsection{Screening for TCs with forbidden transitions}

Using this data set, we perform a high-throughput screening for transparent conductors with forbidden optical transitions at the band edges, which may have been excluded from previous screenings. We assess candidates for both p-type and n-type TCs. Our basic screening methodology is depicted in \autoref{fig:screening-method}(a). We note that the pre-screening steps restrict compounds to those with $E_\mathrm{hull}$ < 0.1 eV/atom as a proxy for stability, and to those with 12 or fewer symmetrically equivalent sites to allow for subsequent hybrid and defect calculations.

\begin{figure*}
    \centering
    \includegraphics[width=\textwidth]{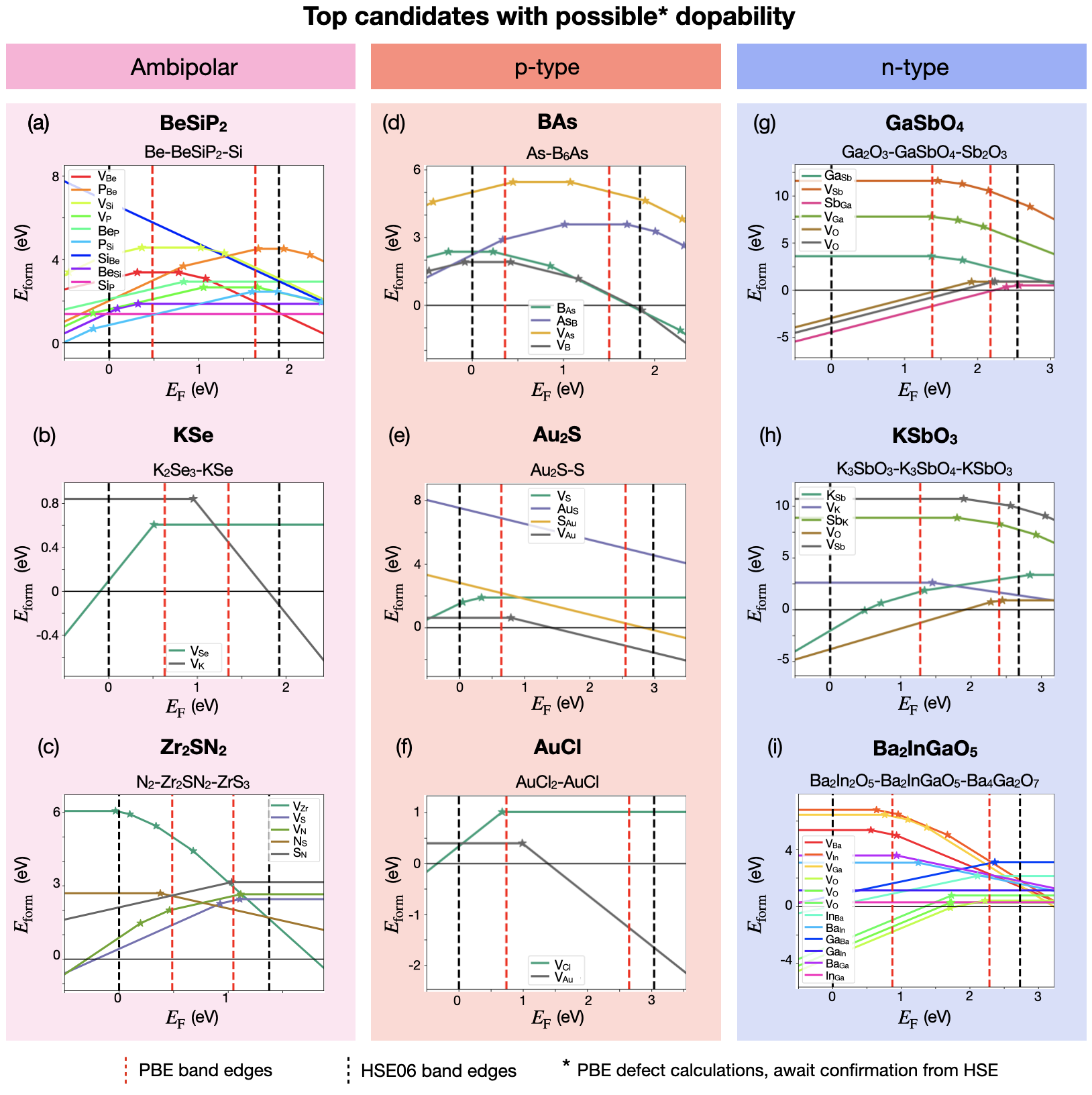}
    \caption{PBE defect formation energy diagrams for a representative set of (a,b,c) candidate ambipolar dopable TCs, (d,e,f) candidate p-type TCs, and (g,h,i) candidate n-type TCs. For each material, only a single representative chemical potential condition is plotted (see SM for other conditions). The KSe diagram plotted here is for \ce{K2Se3}-\ce{KSe} and depicts p-type dopability, whereas n-type dopability is computed at a different chemical potential condition (\ce{K2Se}-\ce{KSe}; see SM). We highlight that these are PBE defect calculations (with an HSE band edge "correction"\cite{broberg2023highthroughput}), and await confirmation of dopability from higher levels of theory.}
    \label{fig:defects}
\end{figure*}


\subsubsection{Screen 1: High-throughput absorption calculations}

In screen 1 we filter compounds based on high-throughput GGA optical absorption calculations and corresponding effective masses at the band edges, as shown in the red-colored screening steps of \autoref{fig:screening-method}(a). Rather than considering the fundamental band gap $E_\mathrm{G}$ or direct band gap $E_\mathrm{G}^\mathrm{d}$, as done in previous screenings for p-type TCs,\cite{hautier2013identification, hautier2014does, bhatia2015highmobility, sarmadian2016easily, williamson2016engineering, shi2017highthroughput,varley2017highthroughput, ha2018computationallydriven, brunin2019transparent} materials are filtered by either their direct allowed gap $E_\mathrm{G}^\mathrm{da}$, the onset of the absorption edge $E_\mathrm{edge}$, or the average absorption coefficient in the visible $\bar{\alpha}_\mathrm{vis}$. Schematics of these descriptors are depicted in \autoref{fig:screening-method}(b). We also prioritize compounds with large $\Delta^\mathrm{d}$ or $\Delta^\mathrm{d}_\mathrm{edge}$, which indicate whether there is a strong presence of optically forbidden transitions at the band edges that could lead to a widening of the absorption edge.

In \autoref{fig:E-vs-meff}(a) and (b), the effective mass $m^*$ is plotted as a function of a GGA energy edge descriptor, either $E_\mathrm{G}^\mathrm{da}$ or $E_\mathrm{edge}$, depending on which value is higher. We restrict our screen to compounds with the GGA energy edge descriptor within the range of 1.5--3.2 eV. Note that for transparency in the visible, absorption edges greater than 3.0 eV are desirable, however PBE can underestimate band gap by a factor of $\sim$1--2 (depending on chemistry and structure)\cite{chan2010efficient} hence the cutoff is reduced by a factor of two. We also include compounds in which the average absorption coefficient $\bar{\alpha}_\mathrm{vis}$ is less than that of reference compound \ce{In2O3} (2.7 $\times$ 10\textsuperscript{3} cm\textsuperscript{-1} using the GGA functional). In this case, even if a material's absorption edge occurs below 1.5 eV, if its total absorption is low enough we still include it in the screen. For the n-type TC screening, we restrict $m^*_\mathrm{e}$ to less than 1, however this tolerance is loosened to $m^*_\mathrm{h}$ < 2 for the p-type TC screening to reflect the much smaller distribution of low $m^*_\mathrm{h}$ than low $m^*_\mathrm{e}$ across materials.\cite{hautier2013identification, woods-robinson2018assessing}

\autoref{fig:E-vs-meff}(a) and (b) plot candidates emerging from screen 1. Pink-colored markers correspond to compounds with large forbidden transitions that would have likely been excluded from previous screens: 579 disperse valence band compounds (plausible p-type TCs) and 790 disperse conduction band compounds (plausible n-type TCs). In total, this amounts to 854 compounds, since most low $m^*_\mathrm{h}$ materials also exhibit low $m^*_\mathrm{e}$. Grey-colored markers correspond to other materials within this range with small or no $\Delta^\mathrm{d}$, and these amount to 5,209 compounds.

\begin{table*}
\centering
\caption{Most promising predicted TC compounds with forbidden optical transitions (see columns $\Delta^\mathrm{d}$ and $\Delta^\mathrm{d}_\mathrm{edge}$), and a summary of their computed optical and electronic properties. See SM for full table.}
\label{tab:hse-forbidden-screen}
\resizebox{\textwidth}{!}{%
\begin{tabular}{ccccccccccccccccc}
\toprule

    \textbf{\begin{tabular}[c]{@{}c@{}}Material\\ ID (mpid)\end{tabular}} 
    &     
    \textbf{Formula} 
    &  
    \textbf{\begin{tabular}[c]{@{}c@{}}Space \\ group\end{tabular}} 
    & \textbf{\begin{tabular}[c]{@{}c@{}} $\bm{E_\mathrm{hull}}$ \\ (eV/at.)\textsuperscript{†}\end{tabular}} 
    &
    \textbf{\begin{tabular}[c]{@{}c@{}} $\bm{E_\mathrm{G}}$ \\ (eV)\textsuperscript{†}\end{tabular}} 
    &
    \textbf{\begin{tabular}[c]{@{}c@{}} $\bm{E_\mathrm{G}}$ \\ (eV)\textsuperscript{‡}\end{tabular}} 
    &
    \textbf{\begin{tabular}[c]{@{}c@{}} $\bm{E_\mathrm{G}^\mathrm{d}}$ \\  (eV)\textsuperscript{‡}\end{tabular}} 
    & 
    \textbf{\begin{tabular}[c]{@{}c@{}} $\bm{E_\mathrm{G}^\mathrm{da}}$ \\ (eV)\textsuperscript{¶}\end{tabular}} 
    & 
    \textbf{\begin{tabular}[c]{@{}c@{}} $\bm{E_\mathrm{edge}}$ \\ (eV)\textsuperscript{¶}\end{tabular}} 
    & 
    \textbf{\begin{tabular}[c]{@{}c@{}} $\bm{\Delta^\mathrm{d}}$\\ (eV)\textsuperscript{†}\end{tabular}} 
    & 
    \textbf{\begin{tabular}[c]{@{}c@{}} $\bm{\Delta^\mathrm{d}_\mathrm{edge}}$\\ (eV)\textsuperscript{¶}\end{tabular}} 
    & 
    \textbf{\begin{tabular}[c]{@{}c@{}} $\bm{\bar{\alpha}_\mathrm{vis}}$\\     (cm\textsuperscript{-1})\textsuperscript{¶}\end{tabular}} 
    &
    \textbf{\begin{tabular}[c]{@{}c@{}} $\bm{m^*_\mathrm{e}}$\textsuperscript{†}\end{tabular}} 
    &
    \textbf{\begin{tabular}[c]{@{}c@{}} $\bm{m^*_\mathrm{h}}$\textsuperscript{†}\end{tabular}} 
    &
    \textbf{\begin{tabular}[c]{@{}c@{}} Doping\\ \small{(defects)}\textsuperscript{†}\end{tabular}} 
    &
    \textbf{\begin{tabular}[c]{@{}c@{}}\#  ICSD \\ entries\end{tabular}} 
    &
 \\

\midrule


    mp-1009085 & \ce{BeSiP2} & $I\bar{4}2d$ & 0.000 & 1.15 & 1.84 & 1.84 & 3.04 & 2.78 & 1.20 & 0.94 & 7.1$\times 10^3$ & 0.35 & 0.47 & p\&n  & 2 \\
    
    mp-9268 & \ce{KSe} & $P\bar{6}2m$ & 0.000 & 0.72 & 1.66 & 2.14 & 2.14 & 3.31 & 0.00 & 1.17 & 4.8$\times 10^3$ &  0.33 & 1.73 & p\&n  & 1\\
    
    mp-11583 & \ce{Zr2SN2} & $P6_3/mmc$ & 0.000 & 0.56 & 1.45 & 2.65 & 3.06 & 2.76 & 0.41 & 0.24 & 8.3$\times 10^3$ &  0.41 & 0.44 & p\&n  & 1\\

    

\midrule

    mp-984718 & \ce{BAs} & $P6_3mc$ & 0.090 & 1.15 & 1.82 & 2.72 & 3.23 & 3.05 & 0.51 & 0.32 & 4.6$\times 10^3$& 0.28 & 0.38 &  p-type  & 0\\
    
    mp-947 & \ce{Au2S} & $Pn\bar{3}m$ & 0.000 & 1.91 & 3.00 & 3.00 & 3.30 & 3.14 & 0.30 & 0.13 & 3.4$\times 10^3$  & 0.42 & 1.55 & p-type  & 2\\
    
    mp-32780 & \ce{AuCl} & $I4_1/amd$ & 0.000 & 1.93 & 3.04 & 3.04 & 3.26 & 3.39 & 0.23 & 0.35 & 1.9$\times 10^3$ & 1.13 & 0.91 & p-type  & 2 \\
    
    
    
\midrule

    mp-1072104 & \ce{GeO2} & $Pnnm$ & 0.006 & 1.40 & 3.23 & 3.23 & 3.38 & 4.02 & 0.15 & 0.79 & 2.2$\times 10^2$ & 0.19 & 1.62 & n-type  &  6 \\
    
    mp-1224786 & \ce{GaSbO4} & $Cmmm$ & 0.000 & 0.80 & 2.47 & 2.47 & 2.83 & 3.34 & 0.36 & 0.87 & 7.7$\times 10^2$ & 0.16 & 1.45 & n-type  &  0 \\
    
    mp-16293 & \ce{KSbO3} & $Fd\bar{3}m$ & 0.045 & 1.12 & 2.58 & 2.89 & 3.35 & 3.32 & 0.46 & 0.43 & 1.0$\times 10^3$ & 0.24 & 0.34 & n-type  &  1 \\
    
    mp-1106089 & \ce{Ba2InGaO5} & $Ima2$ & 0.000 & 1.42 & 2.68 & 2.69 & 3.27 & 3.39 & 0.59 & 0.71 & 1.1$\times 10^3$ & 0.22 & 0.77 & n-type  &  1 \\
    
    mp-1029868 & \ce{Sr5(SiN3)2} & $C12/c1$ & 0.000 & 1.41 & 2.43 & 2.59 & 2.90 & 3.02 & 0.31 & 0.43 & 3.7$\times 10^3$  & 0.34 & 1.20 & n-type  &  0 \\


    
\midrule

    mp-856 & \ce{SnO2} & $P4_2/mnm$ & 0.000 & 0.66 & 2.33 & 2.33 & 3.06 & 3.20 & 0.74 & 0.87 & 1.0$\times 10^3$  & 0.14 & 1.56  & n (ref.) & 42 \\
    
    mp-22598 & \ce{In2O3} & $Ia\bar{3}$ & 0.000 & 0.93 & 2.34 & 2.34 & 2.56 & 3.02 & 0.22 & 0.68 & 2.7$\times 10^3$ & 0.13 & 6.44 & n (ref.) & 18 \\

\bottomrule
\end{tabular}
}
\parbox[t]{\textwidth}{\centering \footnotesize \smallskip
  \textsuperscript{†}GGA calculations.
  \textsuperscript{‡}HSE06 calculations. 
  \textsuperscript{¶}GGA calculations with HSE06 gap correction.
}

\end{table*}

\subsubsection{Screen 2: Medium-throughput HSE calculations}

In screen 2, band gap refinement calculations are applied to the outputs of screen 1 to better approximate direct allowed gap and the absorption coefficient. This approach assumes that the GGA forbidden energy difference $\Delta^\mathrm{d}$ is a sufficient proxy for the difference between HSE $E_\mathrm{G}^\mathrm{d}$ and HSE $E_\mathrm{G}^\mathrm{da}$; however this has not been benchmarked to our knowledge, and $\Delta^\mathrm{d}$ may scale differently depending on functional. Using these HSE shifted energies and spectra, compounds are filtered that fulfill at least one of three criteria as proxies for transparency, as shown in \autoref{fig:screening-method}: $E_\mathrm{G}^\mathrm{da} \geq 2.9$ eV, $E_\mathrm{edge} \geq 2.9$ eV, or $\bar{\alpha}_\mathrm{vis}$ less than that of ITO (2.7 $\times$ 10\textsuperscript{3} cm\textsuperscript{-1}). Outputs are reported in \autoref{fig:E-vs-meff}(c) and (d), yielding 184 previously excluded p-type TC candidates and 243 previously excluded n-type TC candidates.

At this stage, the BPE ratio $\sigma_\mathrm{BPE}$ is also computed as a guideline for whether dopability may be possible. Specifically, BPE energies that lie in the upper quartile of the band gap near the conduction band minimum (CBM; i.e., $\sigma_\mathrm{BPE}$ > 0.75) have been shown to correlate with unlikely p-type dopability, whereas BPE energies that lie in the lower quartile of the band gap near the valence band maximum (VBM; i.e., $\sigma_\mathrm{BPE}$ < 0.25) have been shown to correlate negatively with n-type dopability.\cite{woods-robinson2018assessing} Therefore, we restrict defect calculations in screen 3 to compounds with $\sigma_\mathrm{BPE}$ < 0.75 for p-type candidates and $\sigma_\mathrm{BPE}$ > 0.25 for n-type candidates. Most compounds have mid-gap BPEs so are not excluded from either set. We emphasize that this metric is a guideline that has been demonstrated to correlate with dopability, \textit{not} to predict it, so screens based on BPE should be used with caution.\cite{woods-robinson2018assessing}

\subsubsection{Screen 3: Low-throughput defect and mobility calculations}

In the final screening step, GGA defect formation energy calculations are performed (with HSE VBM and CBM corrections\cite{broberg2023highthroughput}) to assess accessible intrinsic carrier concentrations on $\sim$100 of the most interesting TC candidates to emerge from screen 2. Many of these compounds have unstable defects or intrinsic pinning defects such that they are likely not highly dopable (some may be extrinsically dopable, although this has not been investigated here). We identify a subset of compounds with promising dopability, summarized in \autoref{tab:hse-forbidden-screen} and \autoref{fig:defects}. We will refer to these materials herein as ``candidates'', as each has shown the potential for dopability, but true dopability remains to be confirmed using higher levels of theory and, for instance, hybrid functionals.

First, our predicted p-type dopable TC candidates include \ce{BeSiP2}, \ce{KSe}, \ce{Zr2SN2}, \ce{BAs} (metastable polymorph with space group $P6_3mc$), \ce{Au2S}, and \ce{AuCl}. All except \ce{BAs} are on the convex hull and have been synthesized as bulk materials, while the latter two \ce{Au2S} and \ce{AuCl} have been synthesized as thin films. \ce{Au2S} is a known p-type semiconductor,\cite{ishikawa1995structure} and the dopability of \ce{AuCl} is unknown. \ce{BAs} is a metastable polymorph with space group $P6_3mc$ (its stable cubic polymorph has had recent attention due to its high thermal conductivity\cite{lindsay2013first,kang2018experimental}, and exhibits p-type conductivity\cite{lyons2018impurity}). Compounds \ce{KAlTe2}, \ce{Cd3(BO3)2}, \ce{ScIO}, and \ce{KCuO} are potentially p-type dopable within a smaller window of tolerance; each have been synthesized in bulk but not thin film form. Defect calculations of \ce{ScIO} (and \ce{KCuO}) suggested hole-killing behavior, but only limited chemical potentials were assessed.\cite{wiebeler2020virtual} A few of p-type candidates appear also n-type dopable at various conditions --- \ce{BeSiP2}, \ce{Zr2SN2}, and \ce{KSe} (at \ce{K2Se}-\ce{KSe}, see SM) --- and therefore may be ambipolar dopable semiconductors, however in each case this remains to be confirmed experimentally. We highlight the variety of non-oxide chemistries emerging here; typically forbidden transitions have been studied in oxides, but we demonstrate the importance of looking beyond oxides. In chalcogenides reported here, p-type dopability is limited by anion vacancies.


Compounds in which defects suggest candidate n-type TCs include rutile \ce{GeO2}, \ce{GaSbO4}, \ce{KSbO3}, \ce{Ba2InGaO5}, \ce{Sr5(SiN3)2}, and \ce{LiYS2} (as well as ambipolar candidates \ce{BeSiP2}, \ce{KSe}, and \ce{Zr2SN2}), while \ce{Rb2SnBr6}, \ce{Sr(YS2)2}, and
\ce{GaBiO3} are potentially n-type dopable within a smaller window of tolerance (see SM). Many of these outputs corroborate literature findings. Notably, the two highest-performing, commercially-available n-type TCs --- \ce{In2O3} and \ce{SnO2} --- emerge from the screening at this stage; both have $\Delta^\mathrm{d}$ > 0 eV, and we also include them in \autoref{tab:hse-forbidden-screen} for reference. This is important, as the use of screening parameters from previous studies would have filtered out the best TCs, likely due to their low GGA band gaps (0.66 and 0.93, respectively) and the presence of forbidden transition states at the band edges.\cite{woods-robinson2018assessing} Rutile \ce{GeO2} has been recently studied as an ultra-wide-band-gap material and has been shown to be experimentally ambipolar dopable,\cite{chae2019rutile,chae2021toward} while \ce{GeO2}-derived germanates (e.g., \ce{SrGeO3}) have been explored as n-type TCOs.\cite{mizoguchi2011germanate} Sb-based \ce{GaSbO4} has been explored preliminarily as an n-type TCO.\cite{leung2020synthesis} Perovskite \ce{Rb2SnBr6} has been recently confirmed experimentally as n-type but not yet studied as a TC\cite{ganesan2023influence}, while \ce{SrY2S4} has been grown as a thin film but doping not confirmed. To our knowledge \ce{Ba2GaInO5} has been grown in bulk but not thin film form \cite{didier2014crystal} (a similar compound, brownmillerite \ce{Ba2In2O5}, has shown both n-type and p-type doping and ionic conductivity\cite{fisher1999defect}). All reported oxides are in the main group, corroborating literature consensus on conditions for low effective mass in TCOs;\cite{hautier2014does} we report several non-oxides here, but in each case $m_\mathrm{e}^*$ is not as high as in oxides.

\begin{figure}
    \centering
    \includegraphics[width=0.4\textwidth]{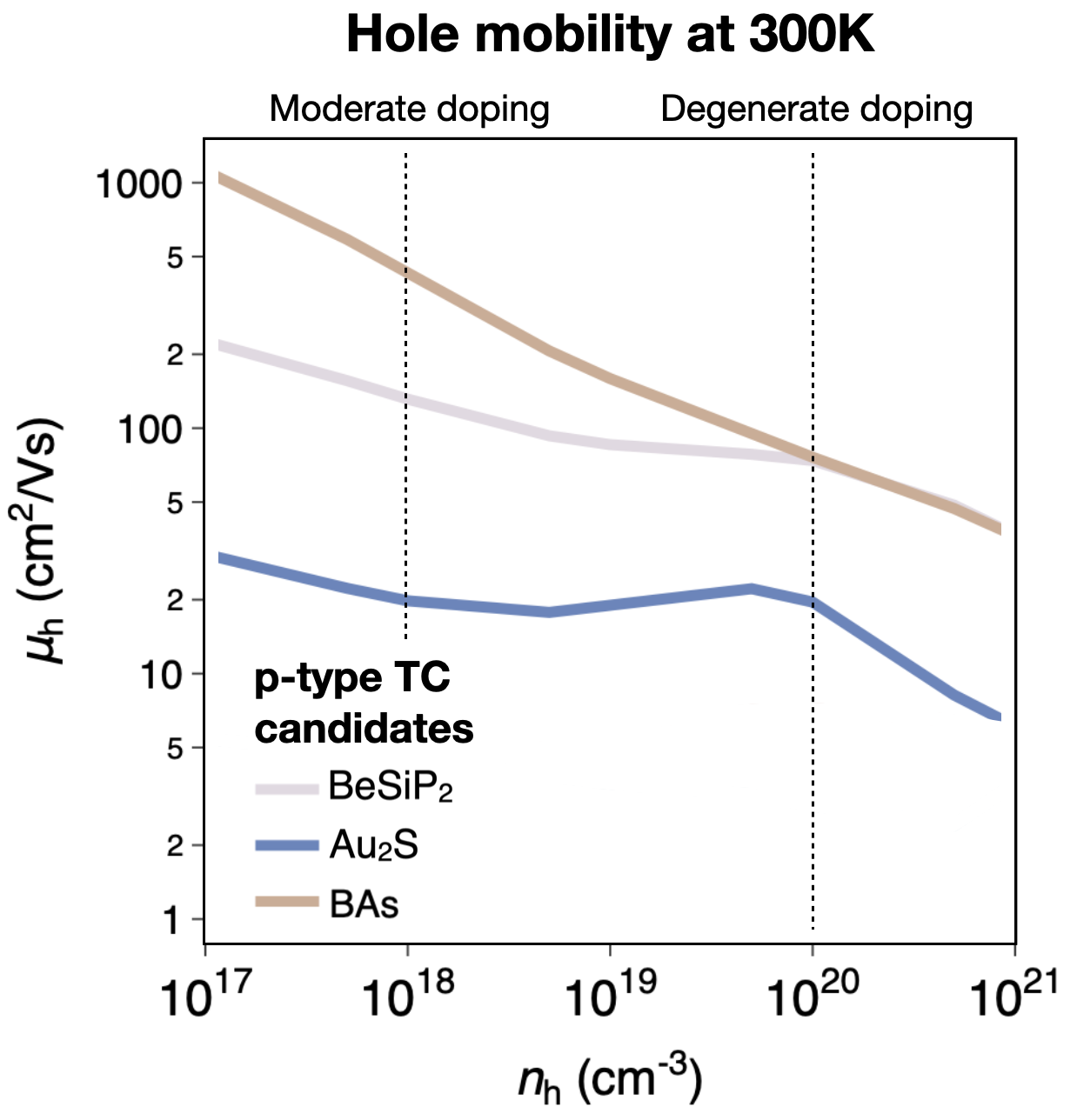}
    \caption{Hole mobility $\mu_\mathrm{h}$ computed with \texttt{amset} code for a representative subset of p-type TC candidates, as a function of doping concentration $n_\mathrm{h}$. See SM for electron mobility calculations.}
    \label{fig:amset}
\end{figure}

To assess charge transport, the mobilities of a few representative candidates are computed using the \texttt{amset} package.\cite{ganose2021efficient} As shown in \autoref{fig:amset}, of these compounds, high hole computed mobilities $\mu_\mathrm{h}$ (greater than 10 cm\textsuperscript{2}/Vs at 300 K at both moderate and degenerate dopings) are exhibited in \ce{BeSiP2},  \ce{BAs}, and \ce{Au2S}, with the former two exhibiting $\mu_\mathrm{h} >$ 100 cm\textsuperscript{2}/Vs. \ce{BeSiP2} also has a computed electron mobility $\mu_\mathrm{e}$ higher than that of standards of \ce{In2O3} and \ce{SnO2}, as shown in the SM. Computed mobilities incorporate polar optical phonon, ionized impurity, and acoustic deformation potential scattering modes. These calculations can be interpreted as an upper limit for scattering in high quality thin films; we do not assess grain-boundary scattering, which is common in thin films.

To confirm dopability in the two most promising p-type TCs to emerge from the screening --- \ce{BeSiP2} and \ce{BAs} --- hybrid defect calculations are performed, as summarized in the SM. These calculations corroborate the PBE defect calculations, suggesting ambipolar dopability in \ce{BeSiP2} (limited by phosphorus vacancies V\textsubscript{P} for p-type doping and beryllium vacancies V\textsubscript{Be} for n-type doping) and p-type dopability \ce{BAs} (limited by boron vacancies, V\textsubscript{B}).

\begin{figure*}
    \centering
    \includegraphics[width=\textwidth]{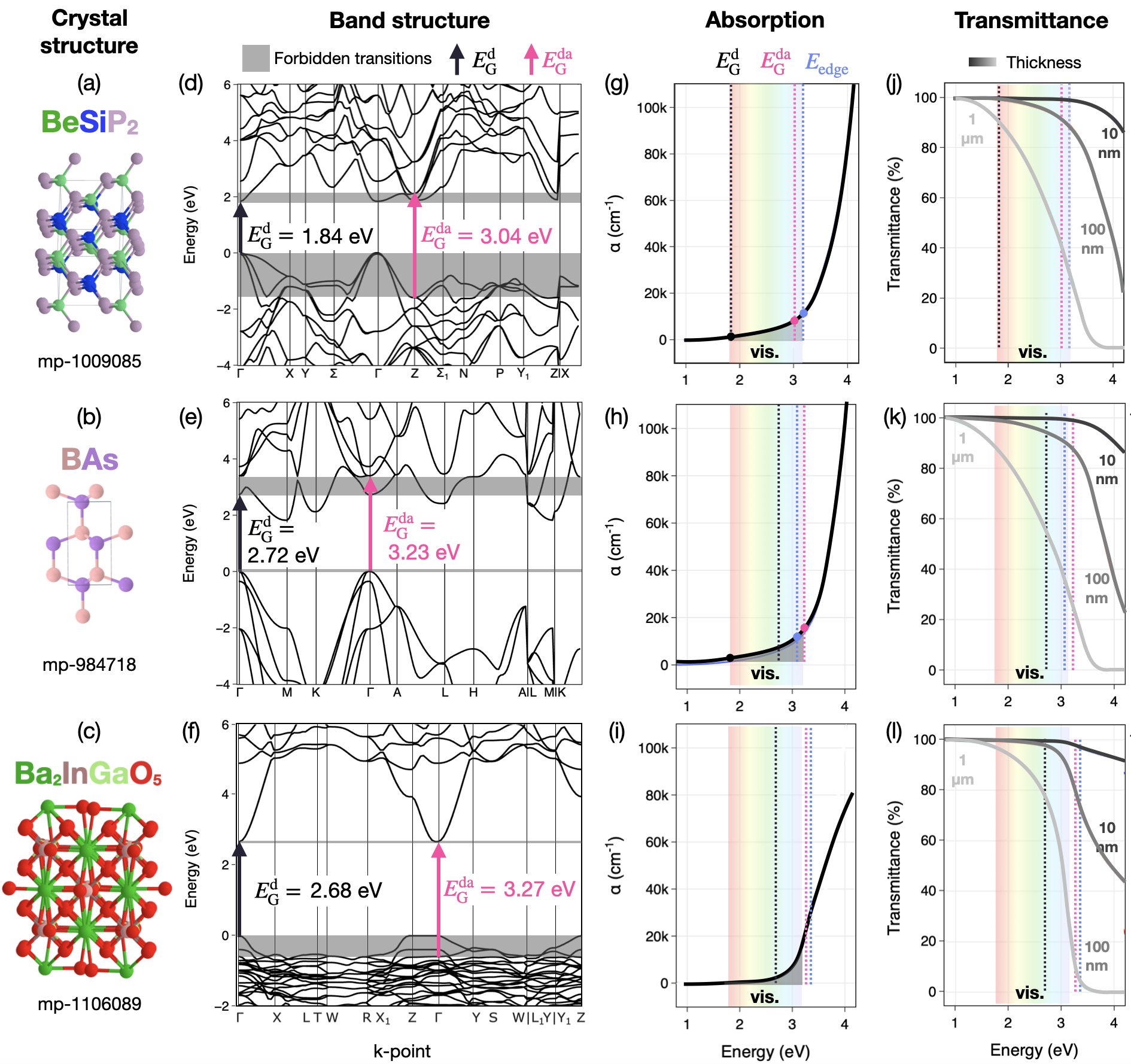}
    \caption{(a--c) Crystal structure, (d--f) HSE-corrected electronic band diagram, (g--i) computed absorption coefficient, and (j--l) computed transmittance for three representative candidates from our screening with forbidden optical transitions: ambipolar-dopable \ce{BeSiP2}, p-type dopable \ce{BAs}, and n-type dopable \ce{Ba2InGaO5}. HSE direct and direct allowed gaps $E_\mathrm{G}^\mathrm{d}$ and $E_\mathrm{G}^\mathrm{da}$ are denoted with black and pink lines, respectively, and in (b) grey shading indicates the region in which optical transitions are forbidden between the VB and CB states. Rainbow shading in (c) and (d) corresponds the visible spectrum (``vis.'').}
    \label{fig:forbidden-Eg-abs-example}
\end{figure*}

\subsection{Identification of candidate TCs with forbidden transitions}

Here, we summarize the optical and electronic properties of the most promising candidates to emerge from GGA defect calculations, as reported in \autoref{tab:hse-forbidden-screen}, and highlight a few examples. Each emerging compound has a GGA gap $E_\mathrm{G}$ below 2 eV, but either an HSE-corrected $E_\mathrm{G}^\mathrm{da}$ or $E_\mathrm{edge}$ greater than 3 eV due to the presence of forbidden transitions or a high $\Delta^\mathrm{d}_\mathrm{edge}$. \ce{Au2S}, \ce{AuCl}, and \ce{GeO2} have HSE gaps $E_\mathrm{G}$ greater than 3 eV so may have emerged from previous screenings but are included here due to their forbidden transitions; \ce{KSe} does not have forbidden transitions, but $\Delta^\mathrm{d}_\mathrm{edge}$ is greater than 1 eV so it is included as well. Compounds in which $\bar{\alpha}_\mathrm{vis}$ is less than that of \ce{In2O3} and likely exhibit a high degree of transparency are \ce{GeO2}, \ce{AuCl}, \ce{KSbO3}, and \ce{Ba2InGaO5}. The former has been investigated in depth, but the latter three would be particularly interesting candidates for follow-up studies.

In \autoref{fig:forbidden-Eg-abs-example} we (a--c) highlight crystal structure and (d--l) optical properties for candidates chalcopyrite \ce{BeSiP2} and wurtzite \ce{BAs}, selected as representative materials since charge transport and hybrid defect calculations have confirmed p-type dopability and mobility, as well as predicted brownmillerite \ce{Ba2InGaO5} as an n-type example. The electronic band structure diagrams of \ce{BeSiP2} (d), \ce{BAs} (e), and \ce{Ba2InGaO5} (f) demonstrate that the direct allowed gap $E_\mathrm{G}^\mathrm{da}$ (pink arrow) is larger than $E_\mathrm{G}^\mathrm{d}$ (black arrow), and the band extrema at $\Gamma$ are very disperse, which leads to low effective masses and high mobilities. The grey shading depict regions in which optical transitions are forbidden. For example, in \ce{Ba2InGaO5} (f) transitions between the upper two VBs along the $\Gamma$-X, $\Gamma$-Y, $\Gamma$-Z paths, as well as the L-T-W path, are forbidden. Thus, the third-highest VB is the highest VB at which transitions between the CBM are allowed, and $E_\mathrm{G}^\mathrm{da,HSE}$ occurs at the $\Gamma$-point at approximately -0.6 eV.  These examples demonstrate three different scenarios in which forbidden transitions can occur. In \ce{Ba2InGaO5} the $E_\mathrm{G}^\mathrm{da}$ and $E_\mathrm{G}^\mathrm{d}$ occur at the same k-point ($\Gamma$) and states are forbidden at the VBM, in \ce{BAs} the $E_\mathrm{G}^\mathrm{da}$ and $E_\mathrm{G}^\mathrm{d}$ occur at the same k-point ($\Gamma$) and states are forbidden at the CBM,  while in \ce{BeSiP2} $E_\mathrm{G}^\mathrm{da}$ occurs at a different k-point (Z) than $E_\mathrm{G}^\mathrm{d}$ ($\Gamma$) and states are forbidden both at the VBM and the CBM.

Panels (g--i) show HSE-corrected absorption coefficient as a function of photon energy, with the edge energy difference $E_\mathrm{edge}$ denoted. In each example material, $E_\mathrm{edge}$ are within a few tens of meV of $E_\mathrm{G}^\mathrm{da}$, although in other candidates this is not necessarily the case (see \autoref{tab:hse-forbidden-screen}, e.g., \ce{In2O3}). Importantly, in both cases $E_\mathrm{edge}$ is at the violet edge of the visible spectrum, which indicates a likelihood of transparency in the visible. Panels (j--k) report transmittance from the Beer–Lambert law (see SM), and the color of the trace corresponds to the thickness of a thin film. As expected, thinner films are more transparent, however the decrease in transparency as thickness increases is material-dependent. For example, although both \ce{BeSiP2} and \ce{Ba2InGaO5} have $>$99\% transmittance for 10 nm thick films, 1 $\mu$m thick films of \ce{Ba2InGaO5} have a $T_\mathrm{vis}$ of $\sim$75\% while $T_\mathrm{vis}$ drops to less than 50\% in \ce{BeSiP2}. however, although \ce{BeSiP2} has $>$99\% transmittance for 10 nm thick films, $T_\mathrm{vis}$ drops to less than 50\% in 1 $\mu$m thick films. Therefore this metric is important when selecting materials for real device applications.

\section{Discussion}

\subsection{Synthesis considerations}

So far we have used simulations to predict properties; the next step for the TC community is to synthesize these materials as thin films and assess their properties experimentally. The final column of \autoref{tab:hse-forbidden-screen} reports the number of experimental ICSD database entries, showing all but four (\ce{BAs}, \ce{Sr5(SiN3)2}, \ce{AlSbO4}, \ce{LiYS2}) have been previously synthesized (although \ce{AlSbO4} has been recently reported\cite{leung2020synthesis}). However in many compounds with ICSD entries, thin films have not yet been grown nor characterized and dopabilty has not been assessed experimentally (e.g., \ce{BeSiP2}, \ce{KSe}, \ce{Ba2InGaO5}, \ce{KAlTe2}, \ce{ScIO}, \ce{KCuO}, etc.).

Some of our candidates have known synthesis challenges, in particular since thin film synthesis is often at non-equilibrium conditions and presents other difficulties. In perovskite oxides \ce{KSbO3} and \ce{GaBiO3} low $m_\mathrm{e}^*$ has been highlighted\cite{hautier2014does} but phase-pure thin film synthesis has proven challenging so doping remains to be confirmed.\cite{homcheunjit2022structural} Non-oxide chalcogenides yield particular synthesis barriers due to decomposition: \ce{KSe} has not been synthesized as a thin film to our knowledge, and \ce{KAlTe2} is likely challenging to synthesize due to oxidation. For candidates that do not have ICSD entries, synthesis may have been attempted but was not successful for various reasons. Wurtzite BAs has been challenging to crystallize and has not yet been synthesized as a thin film to our knowledge; it similar in chemistry to zincblende BP, which was predicted computationally as a p-type TC\cite{varley2017highthroughput} and has since been synthesized as a thin film.\cite{crovetto2022boron}

We acknowledge that some of these candidates are also likely not practical or safe to scale up into device applications. In particular, Be and Be-containing compounds are toxic to humans and the environment,\cite{smith2002toxicological} so although \ce{BeSiP2} has been synthesized it is most likely not a practical TC material. However, since this compound has a common chalcopyrite crystal structure with a small unit cell, understanding the physics behind its large forbidden transition and disperse valence band is demonstrative and could inspire design criteria of other p-type TCs (see \autoref{fig:forbidden-Eg-abs-example}).



\subsection{Challenges and context}

From the set of 18,000 materials with absorption calculations, we have proposed a set of TC candidates with forbidden optical transitions at their band edges and plausible dopability and high mobility. There is a general correlation across all semiconductors that, as the fundamental electronic gap increases, doping becomes more challenging and band edges become less disperse.\cite{zunger2003practical, kittel1986introduction} Previous searches for p-type TCs have endeavored to identify cases in which a single state at the VBM is both disperse over k-space and facilitates transitions which lead to a wide gap. In contrast, by decoupling these two parameters such that allowed transitions do not need to occur at the band edge, our metric could enable better electronic properties while the optical gap is widened. One challenge with this design metric is that localized band edges, which we have shown to correlate with forbidden transitions, tend to lead to \textit{higher} effective masses and therefore lower mobilities. However, we have also demonstrated many candidate TCs with delocalized band edges and forbidden transitions.


The absorption spectra we have computed are first-order, high-throughput approximations, and therefore interpretation of results must consider their limitations. We do not include the effects of spin-orbit coupling, which may influence the orbital character of the band edges or induce spin-forbidden transitions. The IPA accounts only for interband absorption at a fixed k-point, and therefore intraband absorption matrix elements are not considered (phonon-assisted transitions such as indirect gap). This may be sufficient for a first-order approximation since indirect absorption tends to be weak, but in heavily doped TCs, strong free-carrier absorption can arise due to intraband transitions.\cite{pankove1975optical}. Off-stoichiometries and dopants can also introduce shallow defect levels within the gap that reduce optical transparency, and absorption from excitons may also become significant at energies just below the fundamental absorption edge, leading to a reduction of transparency.\cite{pankove1975optical} Despite these limitations, our calculations and data have added information and improved design metrics towards furthering the search for novel TCs.



\section{Conclusion}

In this study, we have described the absorption edge and optical type for $\sim$18,000 semiconductors in the Materials Project database, and we have shared this data publicly on the MPContribs platform. Using a set of descriptors for absorption and orbital character, we have demonstrated correlations between the presence of forbidden optical transitions, localized band edges, and orbital overlap. From this set of materials, we have screened for n-type or p-type TC materials, and propose a set of candidates with forbidden band edge transitions and promising optical and electronic properties such as chalcopyrite \ce{BeSiP2} and wurtzite \ce{BAs}. Notably, high-performance TCs such as ITO emerge from this screening, while being excluded from those based on the fundamental gap alone. Since over half of the set of $\sim$18,000 semiconductors have forbidden optical transitions at their band edges (OT2 or OT4), we recommend that future high-throughput screenings for optical properties use metrics representative of absorption spectra rather than band gap alone.

\section{Acknowledgements}

This work was supported by the U.S. Department of Energy, Office of Science, Office of Basic Energy Sciences, Materials Sciences and Engineering Division under Contract No. DE-AC02-05-CH11231 (Materials Project program KC23MP). R.W.R. was supported by the U.C. Berkeley Chancellor's Fellowship and the National Science Foundation (NSF) Graduate Research Fellowship under Grant No. DGE1106400 and DGE175814. A.M.G. was supported by EPSRC Fellowship EP/T033231/1. We acknowledge compute resources from National Energy Research Scientific Computing Center (NERSC), a DOE Office of Science User Facility. We thank Doug Fabini for helpful discussions, and Ruoxi Yang, Jason Munro, and David Mrdjenovich for fruitful discussion and insights.

\section*{Author contributions}


R.W.R.: Conceptualization, Methodology, Coding, Computational Investigation, Writing - Original Draft, Writing – Review \& Editing, Funding Acquisition;
Y.X.: Computational Investigation (PBE defect formation energy calculations);
J.X.S.: Supervision, Coding;
M.K.H.: Supervision, Methodology, Coding;
N.W.: Computational Investigation (HSE defect formation energy calculations);
M.A.: Funding Acquisition, Resources, Supervision;
A.G.: Computational Investigation (\texttt{amset} calculations);
G.H.: Conceptualization, Supervision, Writing – Review \& Editing;
K.A.P.: Funding Acquisition, Project Administration, Resources, Supervision, Writing – Review \& Editing.


\bibliographystyle{ieeetr}
\bibliography{refs.bib}






\end{document}